\documentclass[twocolumn,pra,aps,showpacs,amsmath,amssymb,floatfix,superscriptaddress]{revtex4-1}

\usepackage{url}
\usepackage{amsmath}
\usepackage{amssymb}
\usepackage{graphicx}
\usepackage{hyperref}
\usepackage{dcolumn}
\usepackage{bm}
\usepackage{xcolor}
\usepackage{siunitx}

\begin{document}
\title{Voltage-controlled magnetic anisotropy in antiferromagnetic MgO-capped MnPt films}
\author{P.-H. Chang}
\author{W. Fang}
\affiliation{Department of Physics and Astronomy and Nebraska Center for Materials and Nanoscience, University of Nebraska-Lincoln, Lincoln, Nebraska 68588, USA}

\author{T. Ozaki}
\affiliation{Institute for Solid State Physics, The University of Tokyo, 5-1-5
Kashiwanoha, Kashiwa, Chiba 277-8581, Japan}

\author{K. D. Belashchenko}

\affiliation{Department of Physics and Astronomy and Nebraska Center for Materials and Nanoscience, University of Nebraska-Lincoln, Lincoln, Nebraska 68588, USA}

\begin{abstract}
The magnetic anisotropy in MgO-capped MnPt films and its voltage control are studied using first-principles calculations. Sharp variation of the magnetic anisotropy with film thickness, especially in the Pt-terminated film, suggests that it may be widely tuned by adjusting the film thickness. In thick films the linear voltage control coefficient is as large as 1.5 and $-0.6$ pJ/Vm for Pt-terminated and Mn-terminated interfaces, respectively. The combination of a widely tunable magnetic anisotropy energy and a large voltage-control coefficient suggest that MgO-capped MnPt films can serve as a versatile platform for magnetic memory and antiferromagnonic applications.

\end{abstract}

\maketitle

\section{Introduction}

Incorporating antiferromagnets (AFM) as active elements in spintronic and magnonic devices \cite{Jungwirth2016,Tserkovnyak2018,Jungwirth2018} could harness their ultrafast dynamics for faster operation and insensitivity to magnetic fields for better scalability and data retention. Although the AFM order parameter can only be switched by very strong magnetic spin-flop fields \cite{Sapozhnik2017,Baldrati2020}, it can also be manipulated by strain \cite{Shick2010,Plekhanov2016,Sapozhnik2017,Liu-piezo2019,Yan2020,Popov2020}, current-induced bulk spin torques \cite{Zelezny2014,Wadley587,Bodnar2018} in AFM of certain symmetries, current-induced interfacial spin torques \cite{Gray2019,Baldrati2020}, and current-induced thermomagnetoelastic effect \cite{Liu2019,Meer2021}. These techniques typically rely on the controlled reorientation of the AFM order parameter between different in-plane orientations. In magnetoelectric AFM, the sign of the order parameter can also be switched by a combination of electric and magnetic fields \cite{He2010}.

Alternatively, interfacial magnetic anisotropy can be tuned by applying an electric field through an electrolyte \cite{Weisheit2007} or, more practically, across a tunnel junction \cite{Duan2008,Ohno2010,Amiri2012}. This technique, called voltage-controlled magnetic anisotropy (VCMA), can be used to facilitate switching in memory devices \cite{memory,memory2,memory3,memory4,Gambardella2020}, control the motion of domain walls \cite{domainwall,dowmainwall2}, and excite and manipulate spin waves in magnonic devices \cite{Hillebrands2014,Verba2014,Parkin2016,SWD2,SWD,SWD4,SWD6,Verba2017,Verba2018,SWD3,Rana2019}.
Although VCMA has been primarily studied at ferromagnetic interfaces, it can also serve as a tuning mechanism in devices with metallic AFM layers \cite{Wang_adma2015,Kioussis2017,Yan2020,Su2020} and even enable coherent AFM domain switching by picosecond voltage pulses \cite{KhaliliVCMA}.

In searching for large VCMA in AFM heterostructures, it is natural to examine AFM materials with heavy elements but relatively small and tunable bulk magnetic anisotropy energy (MAE), which would allow the MAE of a thin film to be engineered to the desired range. One such material is L1$_0$-ordered tetragonal MnPt, which exhibits collinear C-type AFM order, with large 4.3 $\mu_B$ local moments and a N\'eel temperature of 975 K, and can be tuned across spin reorientation transitions (between easy-axis and easy-plane) by off-stoichiometry, temperature variation, and epitaxial strain \cite{Andresen1965,KREN1968,Hama_2007,BUTLER2010,CHANG2018}. It has been shown that MnPt pillars can be reversibly switched between different magnetic states by electric currents \cite{Khalili2020}, and it is compatible with silicon technology \cite{Reiss2020}.

In this paper, we use first-principles calculations to study MAE in MgO/MnPt/MgO films in a wide range of MnPt thicknesses, with both Mn- and Pt-terminated interfaces, and find unusually strong VCMA on the order of 1 pJ/Vm. Based on our results, we propose that MnPt/MgO films can serve as a versatile, tunable platform for antiferromagnonic applications.

\section{Computational methods \label{sec:computational-methods}}

\begin{figure}[htb]
\includegraphics[width=0.9\columnwidth]{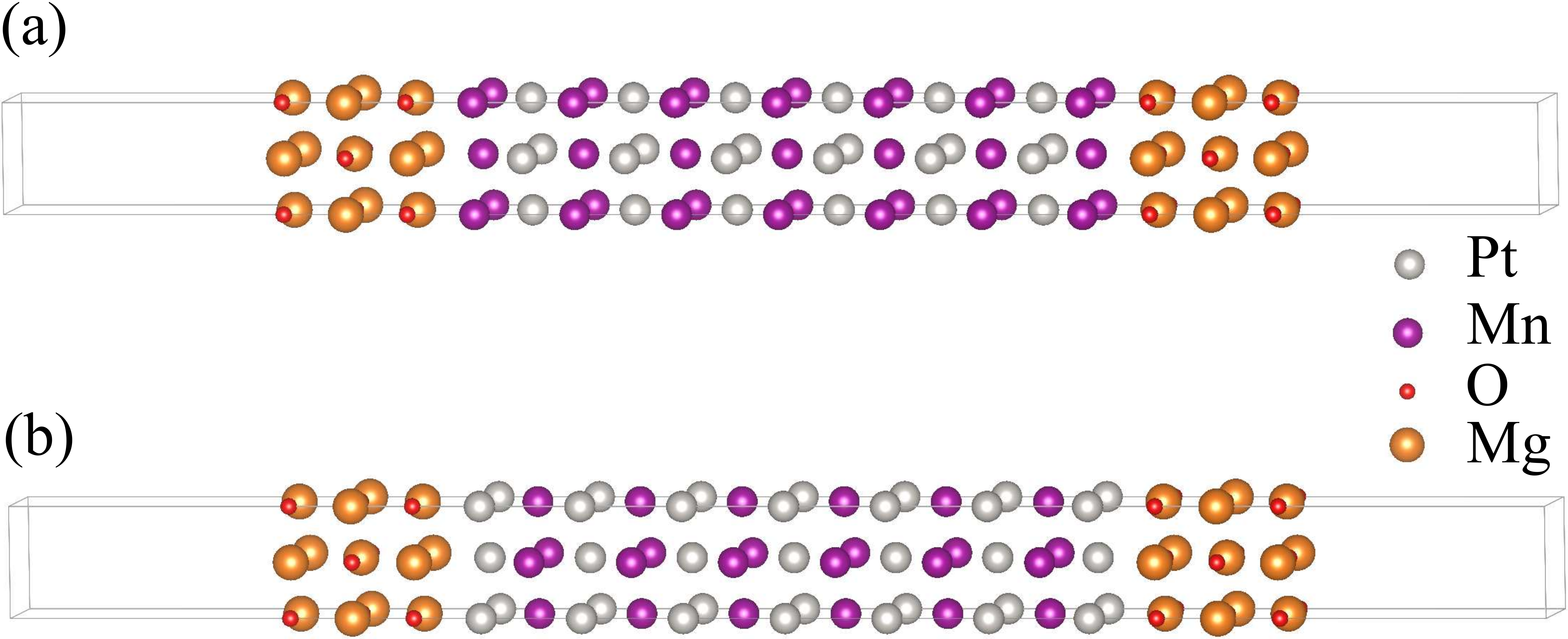}
\caption{Computational setup for (a) Mn-terminated and (b) Pt-terminated MnPt films capped with MgO.}
\label{fig:device}
\end{figure}

We consider L1$_{0}$-MnPt films capped by 3 monolayers (ML) of MgO on each side in a periodic setup with 2 nm of vacuum separating the two MgO layers. Three types of films were considered: with both interfaces terminated by Pt or Mn, and with one interface of each kind; the first two are shown in Fig. \ref{fig:device}. We impose the bulk C-type magnetic structure of MnPt, with staggered magnetic moments in the (001) planes and ferromagnetic spin alignment along the [001] axis.

The structure is optimized using the projector-augmented wave (PAW) method \cite{BLOCHL1994} implemented in the Vienna Ab Initio Simulation Package (VASP) \cite{VASP1999}. The experimental \cite{KREN1968} bulk value of $4.00$ \AA\ is used for the in-plane lattice constant and kept fixed in all calculations. For films with up to 19 ML of MnPt, the atomic coordinates along the out-of-plane axis are relaxed at zero electric field until the forces are less than 1 meV/\AA; the in-plane coordinates are fixed by symmetry. Thicker films are obtained by inserting additional MnPt layers in the middle with the bulk interlayer spacing.

The MAE includes contributions from magnetocrystalline anisotropy (MCA) and magnetostatic dipole-dipole interaction \cite{Szunyogh1995}. We calculate MCA using the OpenMX code with a pseudo-atomic orbital basis set \cite{OZAKI2003,OPENMX,Lejaeghereaad3000}, using the generalized gradient approximation (GGA) \cite{GGA} for exchange and correlation. The charge and spin densities were obtained using a self-consistent calculation without spin-orbit coupling (SOC) and kept fixed in subsequent MCA calculations. At zero field, the MCA was determined as the difference in the total band energy for the configurations with the magnetization aligned in-plane and out-of-plane, with SOC included. The dipole-dipole contribution was calculated by direct real-space summation, which converges absolutely in the two-dimensional film geometry.

The electric field was introduced by inserting an electrostatic dipole layer in the middle of the vacuum region. This setup makes the electric field point outward on one MgO/MnPt interface and inward on the other, and the linear effect of the field on the total MAE of the film is zero. Therefore, the analysis of VCMA requires the contributions of the two interfaces to be separated. This is often done using an approximate site-resolved representation, which replaces MCA by the anisotropy of the SOC energy divided by 2 \cite{ANTROPOV201435}. This representation is acceptable only if MCA is well described by second-order perturbation theory in SOC, which is not always the case. We found that this approximate relation does not hold in MnPt films. Therefore, we use the site-resolved grand canonical potential \cite{Li2013a} $\Omega_i = E_{i}-E_{F} N_{i}$, where $E_{i}$ and $N_i$ are the site-resolved band energy and Mullikan population \cite{Cuadrado2012}:
\begin{align}
E_{i}=\mathop\mathrm{Tr}\sum_{j}\hat\rho_{ij}\hat H_{ji}, \label{eq:bfti-1}\\
N_{i}=\mathop\mathrm{Tr}\sum_{j}\hat S_{ij}\hat \rho_{ji}. \label{eq:Ni-1}
\end{align}
Here $i$, $j$ are site indices, $\hat\rho_{ij}$ is density matrix, $\hat H_{ij}$ the Kohn-Sham Hamiltonian in the real-space representation, and $\hat S_{ij}$ the overlap matrix; the trace is taken over spin and orbital indices. The site-resolved MCA can then be found as the difference between the site-resolved grand potentials corresponding to the in-plane and out-of-plane orientations of the magnetic moments: $K_i=\Omega_i(\parallel)-\Omega_i(\perp)$. This is done as a function of the electric field, and the anisotropy of one interface $K_{int}$ is found by summing up $K_i$ for the sites that are closer to the given interface than the other. Note that Eqs.\ (\ref{eq:bfti-1})-(\ref{eq:Ni-1}) partition off-diagonal terms equally between the two sites, and the result depends on the choice of the basis set. However, because the Hamiltonian is short-ranged in the OpenMX basis set, this ambiguity is immaterial as long as the MnPt layer is not too thin.

\section{Magnetic anisotropy at zero field}\label{sec:MAE}

\begin{figure}[htb]
\includegraphics[width=0.9\columnwidth]{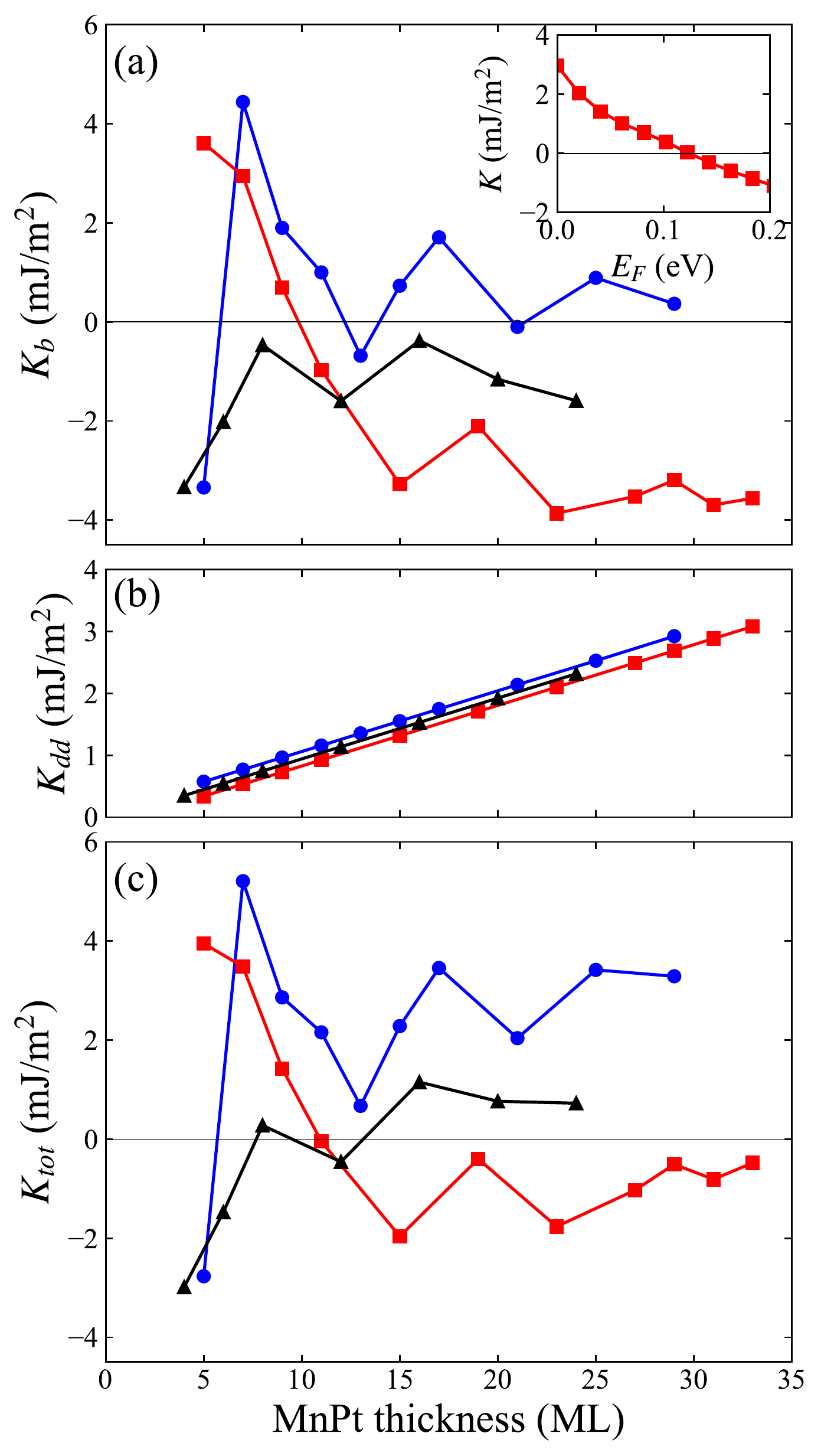}
\caption{Thickness dependence of (a) the band contribution $K_b$, (b) the dipole-dipole contribution $K_{dd}$, and (c) the total MAE $K_{tot}$ of MgO-capped MnPt films with two Mn-terminated interfaces (blue circles), two Pt-terminated (red squares), or one interface of each kind (black triangles). Inset: MCA of the 7-ML-thick Pt-terminated film as a function of the Fermi energy (rigid-band calculation).}
\label{fig:mae-d}
\end{figure}

The magnetocrystalline contribution $K_b$ and the dipole-dipole contribution $K_{dd}$ to MAE as a function of film thickness $d$ are shown in Fig. \ref{fig:mae-d}(a) and \ref{fig:mae-d}(b), and their sum in Fig. \ref{fig:mae-d}(c), for the three types of films.

As the thickness of the film is increased, the $K(d)$ dependence should eventually approach a straight line with a slope equal to the bulk MAE. For the $K_b$ contribution, this asymptotic behavior is only approached in rather thick films, especially if the interfaces are Pt-terminated. This asymptotic slope reflects the small bulk MCA in stoichiometric MnPt \cite{KREN1968,BUTLER2010}. The dipole-dipole contribution $K_{dd}$ depends linearly on thickness for all three termination types, behaving as a positive bulk contribution to MAE that is comparable in magnitude and opposite in sign to the MCA.

\begin{figure}[htb]
\includegraphics[width=0.9\columnwidth]{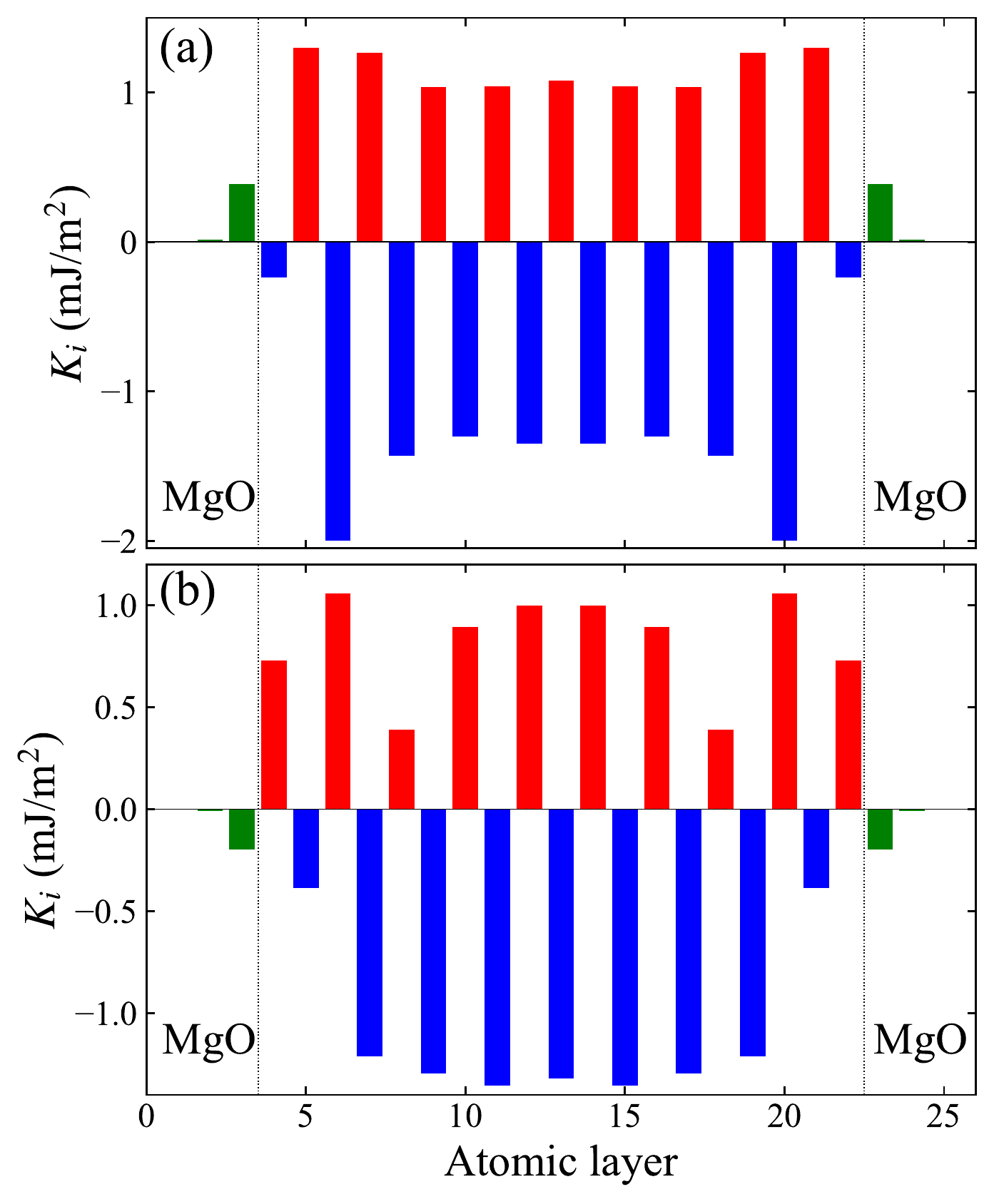}
\caption{Site-resolved MCA in (a) Mn-terminated and (b) Pt-terminated MgO-capped films with 19-ML of MnPt. Blue, red, and green bars show the site-resolved MCA of Mn, Pt, and MgO layers, respectively.}
\label{fig:mca_site}
\end{figure}

Slowly decaying oscillations in the $K_b(d)$ dependence persist up to the largest thicknesses and are likely due to quantum size effects \cite{Bauer2011,Oscillate,Oscillate2}.
Figure \ref{fig:mca_site} shows the site-resolved MCA in 19-ML films with two types of termination. Mn and Pt atoms make large contributions to MCA of opposite sign, but even at 19 ML the site-resolved MCA is not fully converged in the middle of the film. Quantum oscillations may be damped by interface roughness and disorder in an actual sample. As it should, the MCA of a film with one interface of each kind (black line in Fig.\ \ref{fig:mae-d}(a)) asymptotically approaches the average of the films with both interfaces terminated by Pt or Mn, but this also happens at rather large thicknesses.

\begin{figure}[htb]
\includegraphics[width=0.95\columnwidth]{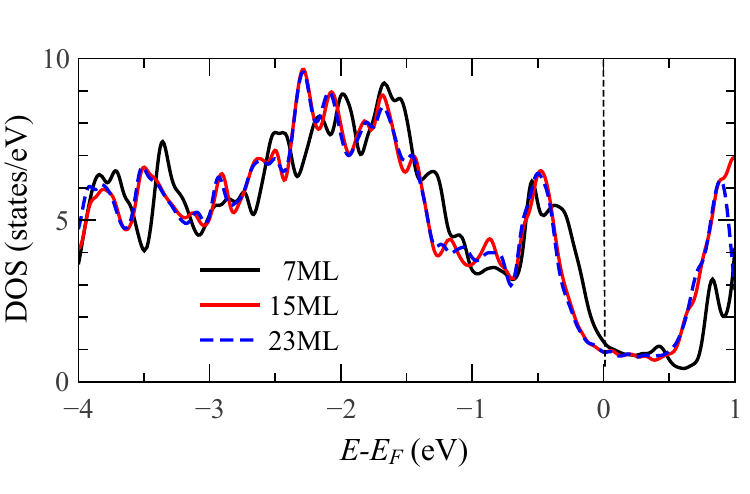}
\caption{Comparison of the contributions from the first four
layers to total DOS among the Pt-terminated films of thicknesses $7$, $15$ and $23$ ML. The dashed line indicates the Fermi level.}
\label{fig:pdos}
\end{figure}

For thicknesses up to 10-15 ML, Fig.\ \ref{fig:mae-d}(a) shows sharp variations in $K_b$. In this region, the two interfaces strongly interact, and it makes no sense to talk about separate bulk and interfacial contributions.
In Mn-terminated films, MCA shows large oscillations, which appear to have the same character as the decaying oscillations at larger thicknesses. However, in Pt-terminated films the MCA declines monotonically from about $4$ to $-3$ mJ/m$^{2}$ as the MnPt thickness increases from 5 to 15 ML.
To understand this decline, we compare the partial density of states (DOS) on the four MLs of MnPt near the Pt-terminated interface for films with $7$, $15$ and $23$ ML of MnPt. As seen in Fig. \ref{fig:pdos}, there is a notable downward shift of about 0.2 eV, between 7 and 15 ML, in the position of the DOS peak right below the Fermi energy.

It is well known that MCA of a metallic system tends to be sensitive to the occupation of the electronic states near the Fermi energy. The inset in Fig. \ref{fig:mae-d} shows the MCA in a 7-ML Pt-terminated film as a function of the Fermi level, calculated in the rigid-band approximation. Raising the Fermi level by 0.2 eV, which corresponds to the band shift between 7 and 15-ML films, reduces the MCA from 3 to $-1$ mJ/m$^{2}$. This is similar to the decline observed in the thickness dependence seen in Fig. \ref{fig:mae-d}.

\section{Voltage-controlled magnetic anisotropy}\label{sec:VCMA}

It is customary to define VCMA with respect to the electric field inside the insulator, which is directly related to gate voltage. Within macroscopic electromagnetism, the field $E_\text{MgO}$ in MgO is reduced by the relative dielectric permittivity $\varepsilon_r$ compared to the field $E_{vac}$ in the vacuum layer.

Static dielectric response in insulators includes comparable contributions from electronic screening and ionic displacements. If the positions of the nuclei are optimized in the presence of the electric field, the electric field in MgO can be obtained, for example, from the magnitudes of the ionic displacements combined with the Born effective charges and force constants. This optimization is, however, computationally challenging. On the other hand, for the Fe/MgO interface it was found \cite{Niranjan2010} that VCMA is insensitive to the field-induced ionic displacements near the interface. This is because VCMA is largely controlled by the areal density of the screening charge at the interface, which is related to the electric induction $D=\varepsilon_0 E_{vac}$ and does not depend on $\varepsilon_r$.

To test this assumption for MnPt/MgO, we first calculate the dielectric constant of bulk MgO using density functional perturbation theory (DFPT) \cite{DFPT-VASP}. The electronic and ionic contributions to $\epsilon_r$ are 3.1 and 6.8, respectively, and the total $\epsilon_r=9.9$ is in excellent agreement with the experimental value \cite{Bartels} of 9.8. Next, we optimized the structure of Mn-terminated and Pt-terminated MgO/MnPt/MgO films with 5 ML of MnPt in an electric field $E_{vac}=1$ V/nm. Using the relative displacement of the Mg and O nuclei, along with the Born effective charges and the force constant calculated from DFPT, we found the electric field in each layer of MgO, which results in the average dielectric constant of 7.4. This value is somewhat smaller than the bulk value, because we are using thin 3-ML layers of MgO. Further, we found that field-induced atomic displacements have a very small effect on VCMA, which is about 3\% for the Mn-terminated and less than 1\% for the Pt-terminated film. Therefore, following Ref.\ \onlinecite{Zhang2017}, in the following we \emph{estimate} the electric field in the MgO layer as $E_\text{MgO}=E_\text{vac}/\varepsilon_r$, where $\varepsilon_r=9.8$ is the experimental value for MgO. The linear VCMA coefficient is then defined as $\beta=\varepsilon_r dK_{int}/dE_{vac}$, where $K_{int}$ is the anisotropy of one interface obtained as explained above after Eqs.\ (\ref{eq:bfti-1})-(\ref{eq:Ni-1}). We have also checked that the dipole-dipole contribution to VCMA is negligible.

The exterior normal to the metallic surface was taken as the positive direction of the electric field. Rather dense $k$-point meshes were needed to converge the linear VCMA coefficient $\beta$: $65\times65$ for Pt-terminated and $35\times35$ for Mn-terminated films, respectively.

\begin{figure}[htb]
\includegraphics[width=0.9\columnwidth]{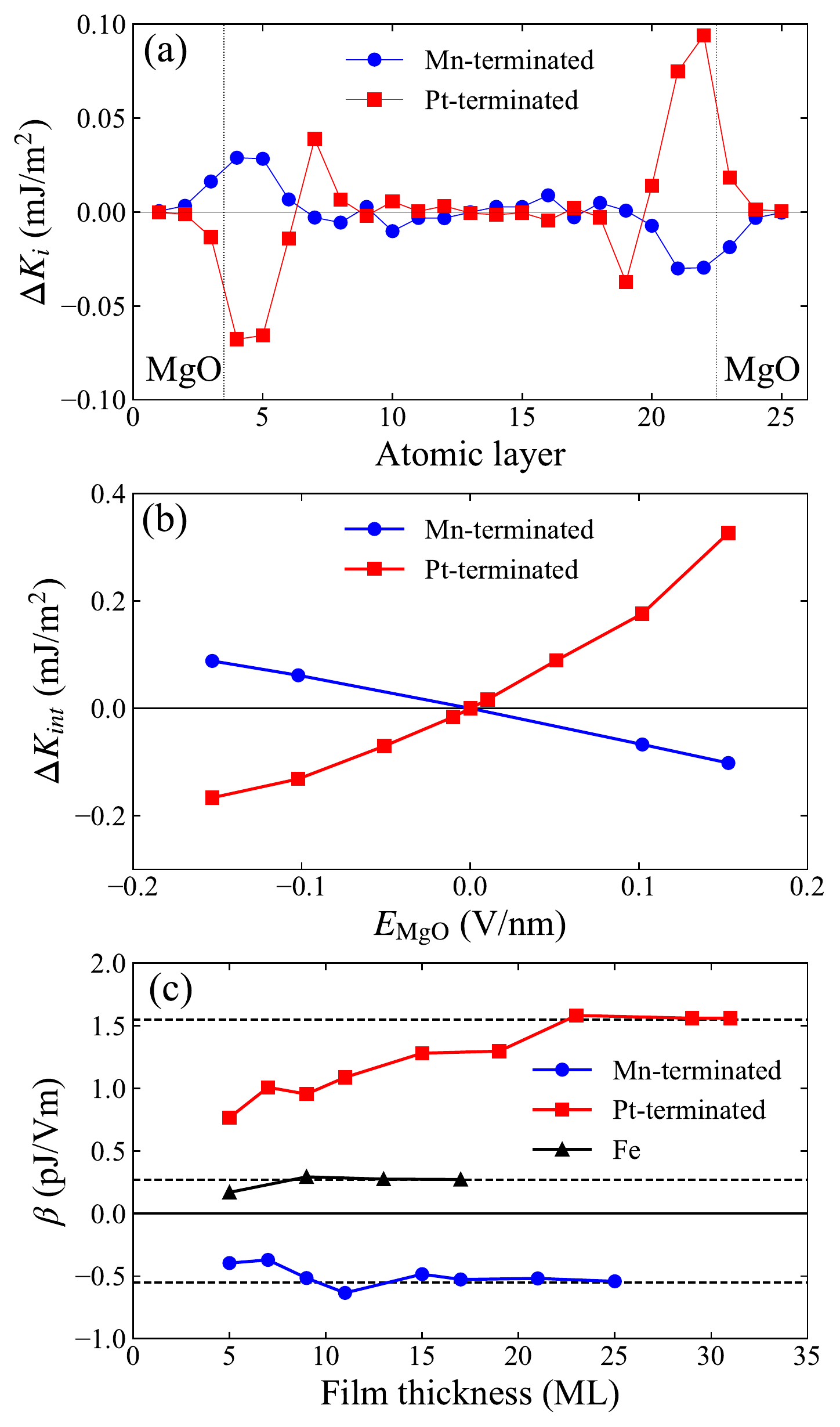}
\caption{(a) Change in the site-resolved MCA induced by $E_{vac}=1.0$ V/nm in Mn-terminated (blue) and Pt-terminated (red) 19-ML films. (b) Induced interfacial anisotropy $\Delta K_{int}$ as a function of the estimated electric field $E_\text{MgO}$ for 15-ML-thick Mn-terminated (blue) and 23-ML-thick Pt-terminated (red) films. (c) Linear VCMA coefficient  $\beta$ in MgO-capped MnPt and Fe films as a function of the thickness of MnPt or Fe.}
\label{fig:betamn}
\end{figure}

Figure \ref{fig:betamn}(a) shows the change $\Delta K_i$ in the site-resolved MAE induced in 19-ML MgO-capped MnPt films by $E_{vac}=1.0$ V/nm. The response in the Pt-terminated film is notably nonlinear in this strong field, which is reflected in deviations from antisymmetry with respect to the middle of the film. We also see that the response has opposite signs for the two interface terminations. The induced anisotropy $\Delta K_i$ is localized within 3 or 4 layers of metal near the interface, which include, for both terminations, two Mn layers closest to the surface and the intervening Pt layers. The MgO layer near the interface also contributes to VCMA. As noted above, the assignment of MAE to atomic sites has an inherent ambiguity on the short length scales corresponding to the range of the atomic orbitals.

Figure \ref{fig:betamn}(b) shows the change in the induced interfacial MAE $\Delta K_{int}$ as a function of the estimated electric field in MgO for 15-ML Mn-terminated and 23-ML Pt-terminated films. These large thicknesses were chosen so that the $\beta$ coefficient is already close to its asymptotic value for the given interface termination. $\Delta K_{int}$ is nearly linear for the Mn-terminated film, but large deviations from linearity are seen for the Pt-terminated one.

The linear VCMA coefficient $\beta$ is plotted as a function of the film thickness in Fig.\ \ref{fig:betamn}(c),
which also includes the results for Fe/MgO films for comparison. In Fe/MgO, $\beta$ is almost constant above 9 ML, and its asymptotic value is 0.26 pJ/Vm. This value agrees well with the result of Ref.\ \onlinecite{Zhang2017} obtained using the charge doping method (0.25 pJ/Vm). The SOC energy method underestimates VCMA at 0.19 pJ/Vm, also in agreement with Ref.\ \onlinecite{Zhang2017}. We note that Ref.\ \cite{Ibrahim2016} obtained a considerably lower value of $\beta$ for the 5-ML Fe/MgO film, which is in part due to the use of the electric field in unrelaxed MgO instead of the experimental $\epsilon$.

The asymptotic $\beta$ coefficient for the Pt-terminated MnPt/MgO interface (1.5 pJ/Vm in thick films) is much larger compared to the Fe/MgO interface. Mn-terminated interfaces have a smaller, but still large, $\beta$ of an opposite sign ($-0.6$ pJ/Vm in thick films). Larger VCMA for the Pt-terminated surface, compared to Mn-terminated, was also found theoretically and experimentally for FePt \cite{Tsujikawa2009a,Miwa2017}.

Different signs of VCMA for Pt- and Mn-terminated MnPt/MgO interfaces can be understood under a simple assumption that interfacial MCA responds in a similar way to the field-induced charge accumulation at the interface as to the uniform shift of the Fermi level in the film. The induced charge density at the interface is $\delta\sigma=\epsilon_0E_{vac}$. Assuming that this charge is concentrated in the interfacial monolayer, we expect a similar effect on MCA from the Fermi level shift $\delta E_F=\delta\sigma/(e N_{int})$, where $N_{int}$ is the partial DOS in that monolayer per unit area, and $e$ is the (negative) electron charge. In this sense, a Fermi level shift $\delta E_F=\epsilon_0 E_{vac}/(e N_{int})$ is ``equivalent'' to the applied field $E_{vac}$. As a crude approximation, we estimate $dK_{int}/dE_F$ as $\frac12 dK_9/dE_F$, where $K_9$ is the total MCA of a MgO-capped MnPt film with 9 ML of MnPt, and the factor $\frac12$ accounts for two interfaces. The result is converted into an estimate of VCMA as $\beta_{e}=\frac12\epsilon_r\epsilon_0 (e N_{int})^{-1} dK_9/dE_F$. The results are shown in Table \ref{tab:fermishift} along with a similar estimate for Fe/MgO. We see that the sign of the estimate $\beta_e$ is correct in all three cases. Given the crudeness of the estimate, even the magnitude of $\beta_e$ can be used as a fair predictor for $\beta$, which may be useful in high-throughput materials design.

\begin{table}[htb]
    \centering
    \begin{tabular}{c|S|S|S|S}
    \hline
    Metal and   & {$dK_{int}/dE_F$} & {$N_{int}$} & {$\beta_e$} & {$\beta$} \\
    termination & {mJ/m$^2$\,eV} & {1/eV\,nm$^{2}$} & {pJ/Vm} & {pJ/Vm} \\
    \hline
    MnPt/Mn & 23.2  & 20.8 & -0.61 & -0.52 \\
    \hline
    MnPt/Pt & -25.4 &  6.8 & 2.02 & 1.08 \\
    \hline
    Fe      &  -7.6 & 34.9 & 0.12 & 0.29 \\
    \hline
    \end{tabular}
    \caption{Estimate of VCMA $\beta_e$ using the Fermi level shift in MgO-capped films with 9 ML of MnPt or Fe (see text).}
    \label{tab:fermishift}
\end{table}

\section{Discussion}

Our results suggest that a thin MgO-capped MnPt film can serve as a versatile platform for antiferromagnonic applications. Both Pt-terminated and Mn-terminated MnPt/MgO interfaces are predicted to have remarkably large VCMA coefficients $|\beta|\sim 1$ pJ/Vm, and the high N\'eel temperature of MnPt is favorable for applications.  A Pt-terminated film appears to be particularly attractive due to its sharp monotonic decline in the total MAE as a function of thickness, from 4 mJ/m$^2$ to $-4$ mJ/m$^2$, in the range between 5 and 15 ML (see Fig.\ \ref{fig:mae-d}). This property may help tune the MAE of the Pt-terminated film to any desired value in this range by adjusting its thickness. Figure\ \ref{fig:mae-d} suggests that such tuning may be possible even if the termination of the film is not strictly controlled. Large sensitivity of the bulk MAE in MnPt to non-stoichiometry, temperature, and strain \cite{KREN1968,Hama_2007,BUTLER2010,CHANG2018} provides additional knobs for tuning the MAE of a film for optimal device performance. Voltage control of anisotropy through local gates can be used to implement spin wave generation, logic, and detection by analogy with ferromagnets \cite{Verba2017,Verba2018,SWD3,Rana2019}, and it may also enable ultrafast switching of AFM order for memory applications \cite{KhaliliVCMA}.

\begin{acknowledgments}

We thank Ilya Krivorotov, Vasyl Tyberkevich, Andrei Slavin, Jia Zhang, and Sai Mu for useful discussions. This work was supported by the Nanoelectronics Research Corporation (NERC), a wholly-owned subsidiary of the Semiconductor Research Corporation (SRC), through the Center for Nanoferroic Devices (CNFD), a SRC-NRI (Nanoelectronics Research Initiative) Center (Task ID 2398.003), and by the National Science Foundation through the Nebraska MRSEC (Grant No. DMR-1420645) and Grants No. DMR-1609776 and DMR-1916275. Calculations were performed utilizing the Holland Computing Center of the University of Nebraska, which receives support from the Nebraska Research Initiative.

\end{acknowledgments}

\end{document}